%% file: Main_1.3.tex
\documentclass[reprint,amsmath,amssymb,aps,prb,superscriptaddress,showpacs,citeautoscript,floatfix]{revtex4-1}
\usepackage{graphicx}
\usepackage{color}
\usepackage{dcolumn}
\usepackage{bm}
\usepackage{amsmath}
\usepackage{amssymb}
\usepackage[]{units}
\usepackage{subcaption}
\usepackage{ragged2e}
\DeclareCaptionJustification{justified}{\justifying}
\usepackage[justification=justified, format=plain]{caption}
\usepackage{booktabs}
\DeclareGraphicsExtensions{.eps,.png,.jpg,.pdf}
\graphicspath{{Figures/}}

\begin{document}

\title[Defects in FePt]{The influence of anti-site defects and stacking faults on the magneto crystalline anisotropy of FePt}

\author{M. Wolloch}
\email{mwo@cms.tuwien.ac.at}
\affiliation{Institute of Applied Physics, Vienna University of Technology, Wiedner Hauptstr. 8-10/134, 1040 Vienna, Austria}
\author{D. Suess}
\affiliation{Institute of Solid State Physics, Vienna University of Technology, Wiedner Hauptstr. 8-10/134, 1040 Vienna, Austria}
\author{P. Mohn}
\affiliation{Institute of Applied Physics, Vienna University of Technology, Wiedner Hauptstr. 8-10/134, 1040 Vienna, Austria}

\begin{abstract}
\input{Abstract.tex}
\end{abstract}


\maketitle

\section{Introduction}
\label{sec:Intro}
\input{Introduction.tex}

\section{Methodological Details}
\label{sec:Method}
\input{Method.tex}

\section{Results and Discussion}
\label{sec:Results}

\subsection{Pristine FePt}
\input{Pristine.tex}

\subsection{Defects in FePt}
\label{sub:defects}
\input{Defects.tex}

\section{Conclusions}
\label{sec:Conclusions}
\input{Conclusion.tex}

\begin{acknowledgments}
\input{Acknowledgements.tex}
\end{acknowledgments}

\end{document}

%% file: Abstract.tex
We present density functional theory (DFT) calculations of the magnetic anisotropy energy (MAE) of FePt, which is of great interest for magnetic recording applications. Our data, and the majority of previously calculated results for perfectly ordered crystals, predict an MAE of \unit[$\sim 3.0$]{meV} per formula unit, which is significantly larger than experimentally measured values. Analyzing the effects of disorder by introducing stacking faults (SFs) and anti site defects (ASDs) in varying concentrations we are able to reconcile calculations with experimental data and show that even a low concentration of ASDs are able to reduce the MAE of FePt considerably. Investigating the effect of exact exchange and electron correlation within the adiabatic-connection dissipation fluctuation theorem in the random phase approximation (ACDFT-RPA) reveals a significantly smaller influence on the MAE. Thus the effect of disorder, and more specifically ASDs, is the crucial factor in explaining the deviation of common DFT calculations of FePt to experimental measurements.

%% file: Introduction.tex
Storage density of hard disc drives (HDDs) have increased over 8 orders of magnitude since their first introduction in the 1950s, peaking in more than 100\% increase per year in the late 1990s and reaching \unit[100]{Gb/in$^2$} in 2002 and, after a period of slower growth, finally \unit[500]{Gb/in$^2$} 2010~\cite{ding:11}. This tremendous achievement was mainly realized through minimization of the read-write head and thinner recording media and reduced grain size. However, to keep storage density growing, grain sizes need to be further reduced and then can be effected by the superparamagnetic limit. Here the magnetic energy stored in a single grain, the product of grain volume $V$ and magnetic anisotropy constant $K_u$, approaches the size of thermal energy $k_\mathrm{B}T$. The thermal stability requirements have thus shifted the focus to materials with very high $K_u$, especially FePt alloys~\cite{weller:13}. While the anisotropy constant of ordered FePt is large enough to allow storage densities of up to \unit[4]{Tb/in$^2$}~\cite{weller:14}, a further problem arises with the limited write fields employed by conventional read-write heads~\cite{richter:12}. Two promising solution to this problem have been proposed, giving a thermal write assist using a laserpuls concentrated by  near field laser optics~\cite{kryder:08}, or exchange spring coupled multilayer media~\cite{suess:05,suess:06}, which reduce the switching field while maintaining good thermal stability. A combination of both approaches is especially promising, for example combining the first order magnetic phase transition of FeRh (see Ref.~\onlinecite{wolloch:16} and Refs. therein) with a small thermal assist to write on extremely hard magnetic FePt alloys~\cite{thiele:04}. To date the highest demonstrated recording density of \unit[1.402]{Tb/in$^2$} was reported in 2015 by using FePt with heat assisted magnetic recording (HAMR)~\cite{ju:15}.

Stoichiometric FePt exists both in the disordered fcc $A1$ phase, and the ordered tetragonal $L1_0$ phase, where Fe and Pt layers are alternating along the $c$ direction. The extremely large magnetic anisotropy energy (MAE) is only found for the ordered phase, which is stable below \unit[$\sim 1300$]{$^\circ$C}~\cite{ding:11}. For magnetic recording, thin films of the material are mainly fabricated by co-sputtering from elemental or alloyed targets~\cite{barmak:05, okamoto:02, thiele:02, thiele:98, wang:16}, and by electron beam evaporation~\cite{farrow:96,kanazawa:00}, but molecular beam epitaxial deposition and other methods are also feasible~\cite{ding:11}. The substrate is mainly MgO(001) and deposition temperature as well as sputtering gas pressure have a large effect on the degree of ordering~\cite{okamoto:02}. Growing extremely highly ordered FePt films with small grains is a challenging endeavor~\cite{ding:11}, but extensive research is performed in order to improve the growth of grains with the easy axis perpendicular to the film plane using special buffer- and seed layers~\cite{shiroyama:17}. This helps with the reduction of the in-plane components of the magnetization, which are a serious noise sources in HAMR.  

\begin{figure}[htbp]
	\centering
    \includegraphics[width=1.0\linewidth]{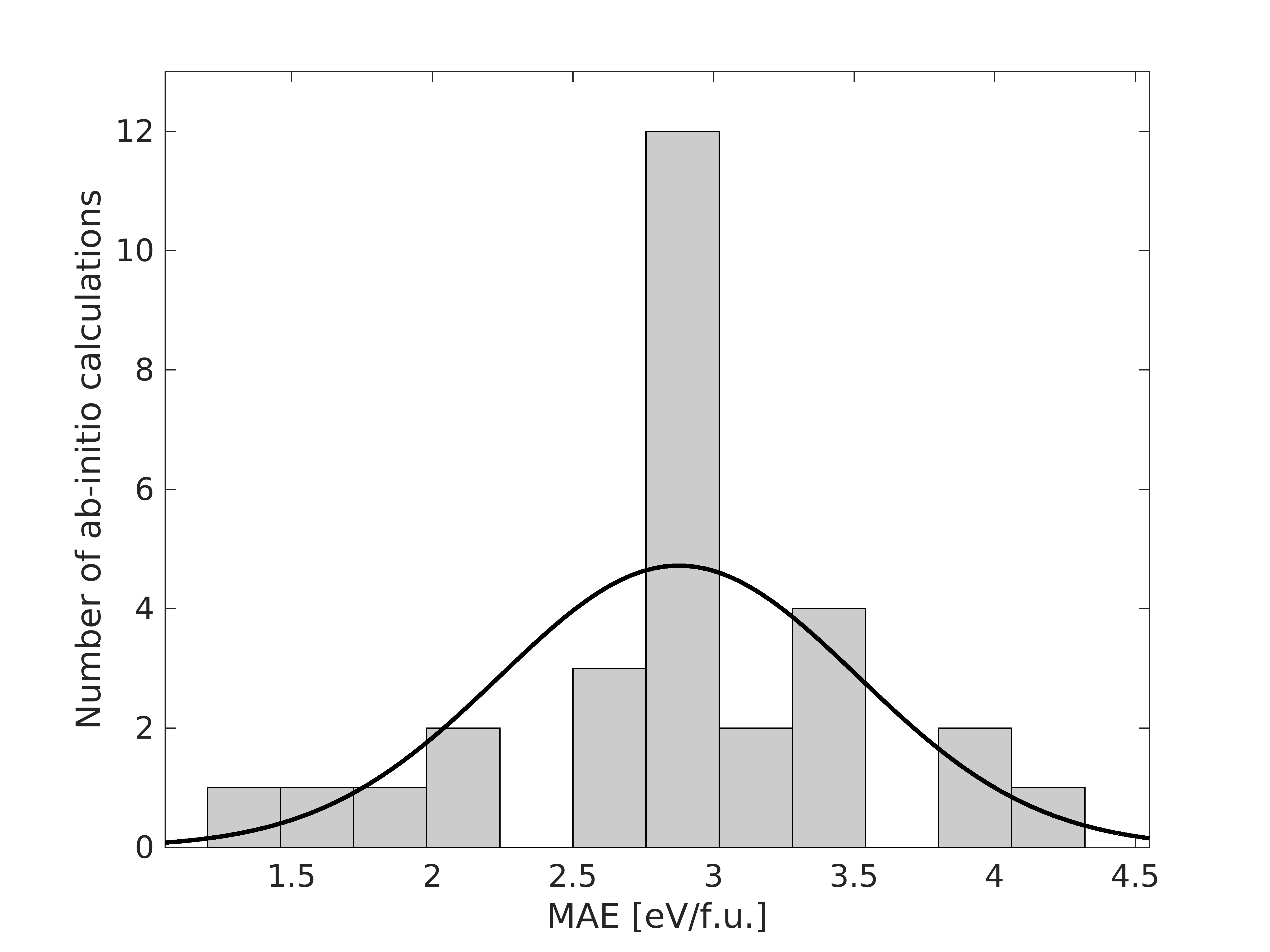}
    \caption{Histogram of 29 previously calculated MAE values of FePt with different DFT codes, computational parameters and exchange and correlation approximations. The black curve is a Gaussian fitted to the data resulting in $\mu=\unit[2.88]{meV/f.u.}$ and $\sigma=\unit[0.64]{meV/f.u.}$. The data are collected from Refs~\onlinecite{aas:11,aas:13,aas:13a,burkert:05,daalderop:91,deak:14,galanakis:00,gruner:13,kabir:15,khan:16,kota:12,luo:14,lyubina:05,ostanin:03,ravindran:01,sakuma:94,shick:03,solovyev:95,staunton:04}.}
	\label{fig:histogram}
\end{figure}

Computationally it is of course much easier to investigate fully ordered FePt using periodic boundary conditions then simulating disordered structures. However, even though FePt has an extremely large MAE it is still only in the meV range per formular unit (f.u.) and calculated by subtracting comparatively large numbers from each other. Additionally, the MAE is a Fermi surface effect and thus very sensitive to the k-point sampling of the Brillouin zone. These effects make accurate calculation challenging and one should not be surprised to find large variations in the results of ab-initio calculations in the literature published in the last decades.
In Fig.~\ref{fig:histogram} we sort 29 previously calculated values for the MAE of FePt, published in 19 different papers~\cite{aas:11,aas:13,aas:13a,burkert:05,daalderop:91,deak:14,galanakis:00,gruner:13,kabir:15,khan:16,kota:12,luo:14,lyubina:05,ostanin:03,ravindran:01,sakuma:94,shick:03,solovyev:95,staunton:04}, in \unit[0.25]{meV} wide bins and fit the data with a Gaussian distribution. All results have been calculated ab-initio with density functional theory (DFT), but involve multiple codes, methods, lattice parameters, and exchange-correlation potentials. Nearly half of the results fall into the bin between \unit[2.75]{meV} and \unit[3.00]{meV}, but the rest of the data are quite scattered, ranging from \unit[1.30]{meV/f.u.} to \unit[4.3]{meV/f.u.}. This leads to an mean of $\mu=\unit[2.88]{meV/f.u.}$ but quite a large standard deviation of $\sigma=\unit[0.64]{meV/f.u.}$.

These values are often compared with the bulk experiments of Ivanov \textit{et al.}~\cite{ivanov:73} from 1973, who used the ballistic throw method and also a vibration magnetometer to measure the magnetic properties of annealed FePt powders, reporting an anisotropy constant of $K_1=\unit[7.0]{MJ/m^3}$, corresponding to \unit[$\sim 1.2$]{meV/f.u.}. Looking at the published computational data, only the results by Shick \textit{et al.}~\cite{shick:03}, and Staunton \textit{et al.}~\cite{staunton:04} are close to this value, at \unit[1.30]{meV/f.u.} and \unit[1.70]{meV/f.u.}, respectively. Ivanov \textit{et al.} argue that their sample is fully ordered, because it exhibits especially high magnetic anisotropy, but take no further measures to actually quantify the degree of order. From more recent experiments on thin films and powders, however (see Sec.~\ref{sub:defects}), we know that full order is not necessary to measure anisotropies of \unit[7.0]{MJ/m$^3$} or higher in FePt, leading us to believe that the sample of Ivanov \textit{et al.} was indeed highly, but not fully ordered~\footnote{Ref.~\onlinecite{ivanov:73} provide their $K_1$ values for room temperature and above. If we extrapolate the data linearly to \unit[0]{K} we arrive at \unit[$\sim 10$]{MJ/m$^3$}, corresponding to \unit[$\sim 1.7$]{meV/f.u.}. This is closer, but still substantially different, from the mean value of the DFT data.}.

In a very recent paper Khan et al. published benchmark calculations for the MAE of fully ordered FePt~\cite{khan:16}. They employ both full-potential linear augmented plane wave (FLAPW) and full-potential Korringa-Kohn-Rostoker (KKR) Green function methods to calculate the MAE within the local density approximation (LDA). The authors of Ref.~\onlinecite{khan:16} took uttermost care in converging their computational parameters and the results of both different DFT methods are in very good agreement with each other. We consider those calculation the most accurate ones published to date. However, since the calculated MEA of \unit[$\sim 3.0$]{meV} is still about twice as large as the reported experimental bulk value of Ref.~\onlinecite{ivanov:73}, Khan et al. conclude that many body effects beyond the LDA are playing a decisive role for the MAE of FePt. In the present paper we will argue that the benchmark calculations of Ref.~\onlinecite{khan:16} and the majority of other computations (see Fig.~\ref{fig:histogram}) are indeed correct for ideally ordered FePt, but that experiments always measure somewhat disordered structures and thus defects must be explicitly considered to reconcile calculations with experiment.

%% file: Method.tex
We have performed spin polarized DFT computations employing the Vienna Ab-Initio Simulation Package \textit{VASP}~\cite{kresse1993,kresse1994a,kresse1996a,kresse1996b} version 5.4.1, using the Projector Augmented-Wave (PAW) method~\cite{bloechl1994,kresse:98}. The plane wave energy cutoff was chosen to be \unit[900]{eV}, which is more than 230\% (300\%) higher than the recommended value for the Fe (Pt) PAW potentials (set of 2003) which treat the  $3s$, $3p$, $3d$, and $4s$ ($5s$, $5p$, $5d$, and $6s$) electrons (32 per FePt pair) as valence. We sample the Brillouine zone with generalized Monkhorst-Pack grids as described by Wisesa et al., finding a significantly quicker convergence with their server generated grids than for those generated by the \textit{VASP} routines~\cite{wisesa:16}. Unless otherwise noted their parameter $r_\mathrm{min}$, which describes the distance between lattice points on the real-space superlattice and increases k-mesh density if increased, was set to \unit[65]{\AA}.  The chosen energy cutoff might seem large, but only with this cutoff we can achieve a total energy convergence of less than \unit[0.1]{meV}, and thus properly quantify the MCA.
To approximate the effects of exchange and correlation the generalized gradient correction (GGA) as parametrized by Perdew, Burke, and Ernzerhof (PBE) has been used~\cite{PBE}.
To ensure accurate forces during relaxations we use an additional superfine fast Fourier transform (FFT) grid for the evaluation of the augmentation charges and a smearing of \unit[$\leq0.1$]{eV} according to Methfessel and Paxton~\cite{methfessel:89} (first order). For total energy calculations the tetrahedron method with Bl\"ochl corrections has been used~\cite{bloechl:94}. In all total energy GGA calculations we explicitly account for non spherical contributions of the gradient corrections inside the PAW spheres. Electronic relaxations are converged to \unit[$10^{-5}$]{meV}, while forces in ionic relaxations where converged to \unit[$\leq1$]{meV/\AA}.
For all calculations of the MAE we turned all symmetry options off explicitly and subtracted the total energy values of hard and easy axis orientation of the magnetic moments. The easy axis is clearly the [001] direction with a hard plane orthogonal to it in which the energy difference between orientations is in the $\mu$eV range. We chose the [110] direction of the $L1_0$ unit cell as the hard axis (corresponding to the [001] direction if distorted fcc unit cell is used). 

%% file: Pristine.tex
The high MAE of FePt in the L$1_0$ phase is mainly due to the large spin orbit coupling in the Pt atoms. They show magnetic moments induced by the Fe $3d$ orbitals, and the $d$ orbital of both species hybridize with each other. Detailed discussions about the origin and nature of the large MAE can be found in the literature~\cite{daalderop:91,solovyev:95,kota:12,khan:16} and will not be discussed here further.

In contrast to Ref.~\onlinecite{khan:16} we use the PBE functional, which belongs to the class of GGAs, instead of the LDA. This choice was made due to the better equilibrium volume obtained by PBE of \unit[28.02]{\AA$^3$}, which at $+2\%$, is much closer to the experimental value of \unit[27.5]{\AA$^3$}~\cite{lyubina:05} than LDA at \unit[24.55]{\AA$^3$} ($-11\%$). Since we need to relax our structure once we introduce defects, getting better volumes and forces is even more important. Additionally, at the smaller equilibrium volume given by the LDA the local magnetic moments of the Fe and Pt atoms would be reduced by about 15\% and 10\%, respectively.

Relaxing $L1_0$ FePt with PBE yields lattice parameters $a=\unit[2.7287]{\AA}$ and $c=\unit[3.7629]{\AA}$. This leads to a $\nicefrac{c}{a}$ ratio of 1.38, about 1.5\% larger than the experimental value of 1.36~\cite{lyubina:05}. While this difference will influence the MAE somewhat, the effect of the $\nicefrac{c}{a}$ ratio is considered to be small compared to disorder in the sample\cite{ostanin:03}.

For PBE we calculate an MAE of \unit[2.74]{meV/f.u.}, corresponding to an anisotropy constant $K_u$ of \unit[15.7]{MJ/m$^3$}. Increasing the total number of k-points in the Brillouin zone by $\sim 55\%$ from 7317 to 11340 (via adjusting $r_\mathrm{min}$ to \unit[75]{\AA}) does not change this value. This result compares well to the mean value of previously published results of Fig.~\ref{fig:histogram}, and is in excellent agreement with the benchmark calculations of Ref.~\onlinecite{khan:16}, which reports a value of \unit[2.73]{meV/f.u.}~for PBE. The angular dependence of the anisotropy Energy $E$ can be fitted to $E(\theta)-E(0)=K_1 \sin^2(\theta) + K_2 \sin^4(\theta)$, where we find that $K_1$ with \unit[2.67]{meV} is more than one order of magnitude larger than $K_2$ with \unit[0.13]{meV}. Employing the LDA functional (for the same PBE-relaxed unit cell) yields a somewhat larger value of \unit[3.11]{meV/f.u.} (which is virtually unchanged if we calculate the MAE for the slightly smaller lattice parameters used in Ref.~\onlinecite{khan:16}), again in good agreement with the benchmark calculations (reporting values from 2.85 to \unit[3.12]{meV/f.u.}, depending on method and code)~\cite{khan:16}. The local magnetic moments as well as the orbital moments calculated with PBE, LDA, and LDA+U can be found in Tab.~\ref{tab:moments}.

\begin{table}[hbt]
\caption{Local spin moments ($m^\mathrm{loc}$) and orbital magnetic moments ($m^\mathrm{orb}$) for FePt. $\|$ and $\bot$ mark orientation of the spin moments parallel and normal to the z-axis, respectively.}
\label{tab:moments}
\begin{ruledtabular}
\begin{tabular}{lcccccc}
 \noalign{\vskip 1mm}
 & \multicolumn{2}{c}{$m^\mathrm{loc}$ [$\mu_\mathrm{B}$]} & \multicolumn{4}{c}{$m^\mathrm{orb}$ [$\mu_\mathrm{B}$]} \\
  \noalign{\vskip 1mm}
 & Fe & Pt & Fe$_\|$ & Pt$_\|$ & Fe$_\bot$ & Pt$_\bot$ \\
 \noalign{\vskip 1mm}
\hline
\noalign{\vskip 1mm}
PBE	  & 2.83 & 0.39 & 0.056 & 0.044 & 0.052 & 0.057 \\
LDA   & 2.69 & 0.37 & 0.058 & 0.043 & 0.052 & 0.055 \\
LDA+U & 2.83 & 0.36 & 0.056 & 0.043 & 0.054 & 0.055 \\
\end{tabular}
\end{ruledtabular}
\end{table}

\subsubsection{Analyzing correlation effects}
\label{subsub:RPA}

Although we are able to reproduce the reference values of Khan et al.~very well with less computational effort, we fail to reproduce the LDA+U results published previously by Shick and Myrasov~\cite{shick:03}. Using their lattice parameters and values for both U and J, we calculate a MAE of \unit[2.79]{meV/f.u.}, which is about \unit[0.32]{meV} less than our result with LDA for their lattice parameters, but still about twice as large as their published result of \unit[1.3]{meV/f.u.}. Switching from the rotationally invariant LDA+U flavor of Liechtenstein et al.~\cite{liechtenstein:95}, to the simplified version by Dudarev et al.~\cite{dudarev:98}, did not change our result significantly.

To investigate correlation effects further we employ the adiabatic connection fluctuation-dissipation theorem in the random phase approximation (ACFDT-RPA), as implemented in the VASP package~\cite{harl:08,harl:09,harl:10}. Here the correlation energy is computed via the plasmon fluctuation equation by calculating the independent particle response functions using occupied and unoccupied states. The exchange energy is calculated by Hartree-Fock theory. Both contributions are calculated non self-consistently using DFT orbitals and are added to Hartree, kinetic, and Ewald energies to obtain the total ACFDT-RPA energy. As prudent for metallic systems, we neglected long wavelength contributions~\cite{harl:10}.

Due to the huge computational effort needed for such calculations, we where not able to perform them with the same accuracy as our other MAE calculations. Furthermore the integral over $\omega$ in the plasmon equation has to be solved numerically using a fixed number of sampling points $N_\omega$. Fortunately convergence with respect to $N_\omega$ (which can be troublesome for metals) was quick for FePt, and the necessary accuracy (\unit[$\sim 0.1$]{meV}) was already obtained for $N_\omega=10$. Calculating the RPA energy for significantly more than $\sim 1000$ k-points proved impossible with our computational resources, so we reduced the $r_\mathrm{min}$ parameter to maximally \unit[34]{\AA}, resulting in 1088 k-points in the full Brillouin zone. The plane wave cutoff was also reduced to \unit[600]{eV}, which then leads to an MAE of \unit[3.1]{meV/f.u.} on the PBE level, about 10\% higher than for our converged computational parameters.

\begin{figure}[htbp]
	\centering
    \includegraphics[width=\linewidth]{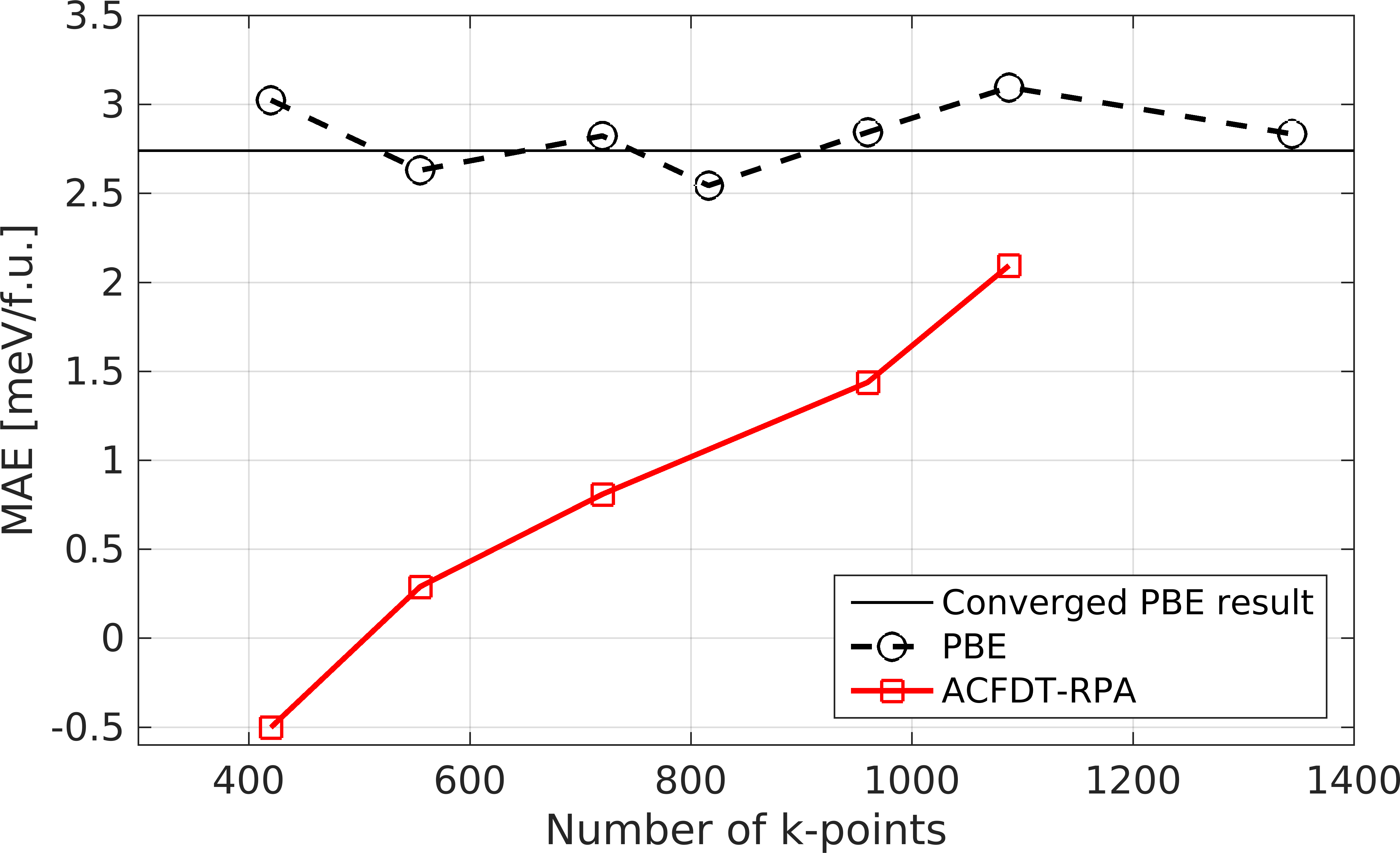}
    \caption{MAE calculated with on the ACFDT-RPA level with respect to the number of k-points in the full Brillouin zone. The PBE results for an energy cutoff of \unit[600]{eV} are also plotted for comparison. The MAE increases monotonically with increased number of k-points and should be higher than \unit[2.0]{meV/f.u.}.}
	\label{fig:RPA_MAE}
\end{figure}

As can be seen from Fig.~\ref{fig:RPA_MAE}, the ACFDT-RPA calculations are not at all converged at 1088 k-points, much in contrast to standard DFT, which is qualitatively correct even for comparatively low k-mesh densities and for an energy cutoff of \unit[600]{eV}. For example 420 k-points are enough to approach the converged value of the MAE within $\sim 10\%$ for PBE, while on the RPA level not even the easy axis is correctly predicted. At least we are able to show a clear trend in our data, where the MAE increases monotonically with the number of k-points for ACFDT-RPA calculations. Of course it is not possible to predict at what value of the MAE the ACFDT-RPA results will converge from the trend at low k-mesh densities, but we are fairly confident that the data is sufficient to predict a higher value than \unit[2.0]{meV/f.u.}.

Although we could not converge our ACFDT-RPA calculations, the trend we observe make us confident that correlation effects alone are not able to reduce the MAE of FePt by a factor of two. In the following section we will show that disorder is able to reconcile experiment and calculations much more satisfyingly than a high level treatment of exchange and correlation.

%% file: Defects.tex
\begin{figure}[htbp]
	\centering
    \includegraphics[width=\linewidth]{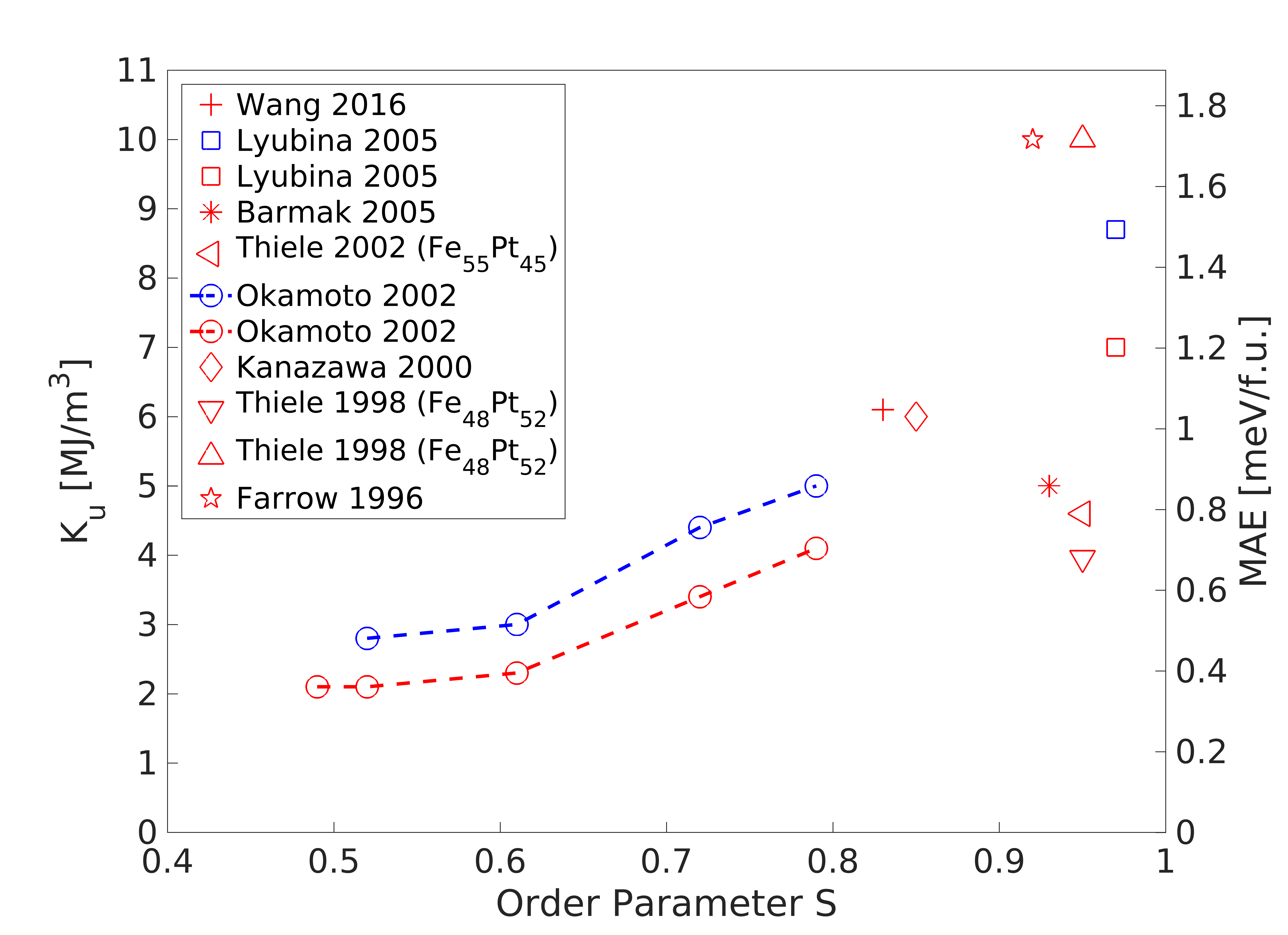}
    \caption{Experimental anisotropy constant $K_u$ in MJ/m$^3$ and plotted over the long range order parameter $S$. The right axis is a conversion to the MAE in meV. Data is taken from Refs.~\cite{wang:16,lyubina:05,barmak:05,thiele:02,okamoto:02,kanazawa:00,thiele:98,farrow:96} as indicated in the legend. Red symbols are measurements at room temperature, while blue symbols stand for low temperature.}
	\label{fig:Exp}
\end{figure}

As discussed in section~\ref{sec:Intro}, experimental measurements of the MAE of FePt are always performed for a somewhat disordered crystal. Disorder in a crystal can be quantified by the long range order parameter $S$. In the case of a stoichiometric FePt crystal, the fractions of Fe and Pt atoms sitting on their correct respective lattice sites must be equal ($r_\mathrm{Fe} = r_\mathrm{Pt} = r$), thus the equation for $S$ reduces to
\begin{equation}
\label{equ:order}
S= r_\mathrm{Fe}+r_\mathrm{Pt}-1 = 2 r-1 \quad.
\end{equation}
For a totally disordered crystal $S=0$, as each atom has 50\% probability to sit on its preferred lattice site, while $S=1$ is achieved for perfect order.
\footnote{Sometimes also the short range order parameter $\eta$ is used in this context. $S$ and $\eta$ are connected by the relation $\eta = 2S -1$, and $0.5 \leq \eta \leq 1$.}
Experimentally the order parameter usually is estimated by the relative strength of integrated X-ray diffraction peaks $I_{001}$ and $I_{002}$ according to the formula
\begin{equation}
\label{equ:exp_order}
S^2 = \frac{\left(I_{001} / I_{002} \right)_\mathrm{meas}}{\left(I_{001} / I_{002} \right)_\mathrm{calc}^{S=1}} \quad ,
\end{equation}
where the numerator consists of the measured values and the denominator uses calculated intensities for perfect order, assuming atomic scattering factors, Debye-Waller correction, Lorentz polarizations factors and structure factors~\cite{wang:16}. However, in a recent investigation of a multi-grain FePt nanoparticle by 3D atomic electron tomography, it was observed that L$1_0$ order might be wrongly attributed in standard 2D methods due to overlapping L$1_2$ grains, although this seems unlikely in highly stoichiometric samples~\cite{yang:17}.
In Fig.~\ref{fig:Exp} we have plotted several experimentally determined values for the magnetic anisotropy constant $K_u=K_1+K_2$. Values are given in MJ/m$^3$ and have also been converted to meV/f.u., for easier comparison to calculations. Most measurements have been performed at room temperature (RT shown with red symbols) but Okamoto \textit{et al.}~\cite{okamoto:02} and Lyubina \textit{et al.}\cite{lyubina:05} have also provided low temperature measurements at 10 and \unit[5]{K}, respectively (LT shown with blue symbols). From their data we see that the MAE is reduced by $\sim 20\%$ to 30\% at RT compared to LT.
More generally, the temperature dependence of the first order anisotropy constant $K_1$ is coupled to the temperature dependence of the magnetization $M_S$ to approximately second power, $K_1(T)/K_1(0)=\left(M_S(T)/M_S(0)\right)^2$, as measured by Refs.~\onlinecite{thiele:02,okamoto:02}, and calculated by Refs.~\onlinecite{staunton:04,mryasov:05,deak:14,kobayashi:16}.
From Fig.~\ref{fig:Exp} we also see that the spread of values for high order parameters is quite large, an effect which can be explained in part by the different compositions of the samples (see legend of Fig.~\ref{fig:Exp}), but also indicates the difficulties in accurately measuring such a large anisotropy with usual laboratory fields. For example Thiele \textit{et al.}~\cite{thiele:98} give two values for the MAE of the same sample, once measured by torque magnetometry (\unit[3.96]{MJ/m$^3$}, $\triangledown$ in Fig~\ref{fig:Exp}), and once deduced from saturation magnetization and dipolar length measurements (\unit[10]{MJ/m$^3$}, $\vartriangle$ in Fig~\ref{fig:Exp}), which differ by more than 100\%.

\begin{figure}[htbp]
	\centering
    \includegraphics[width=\linewidth]{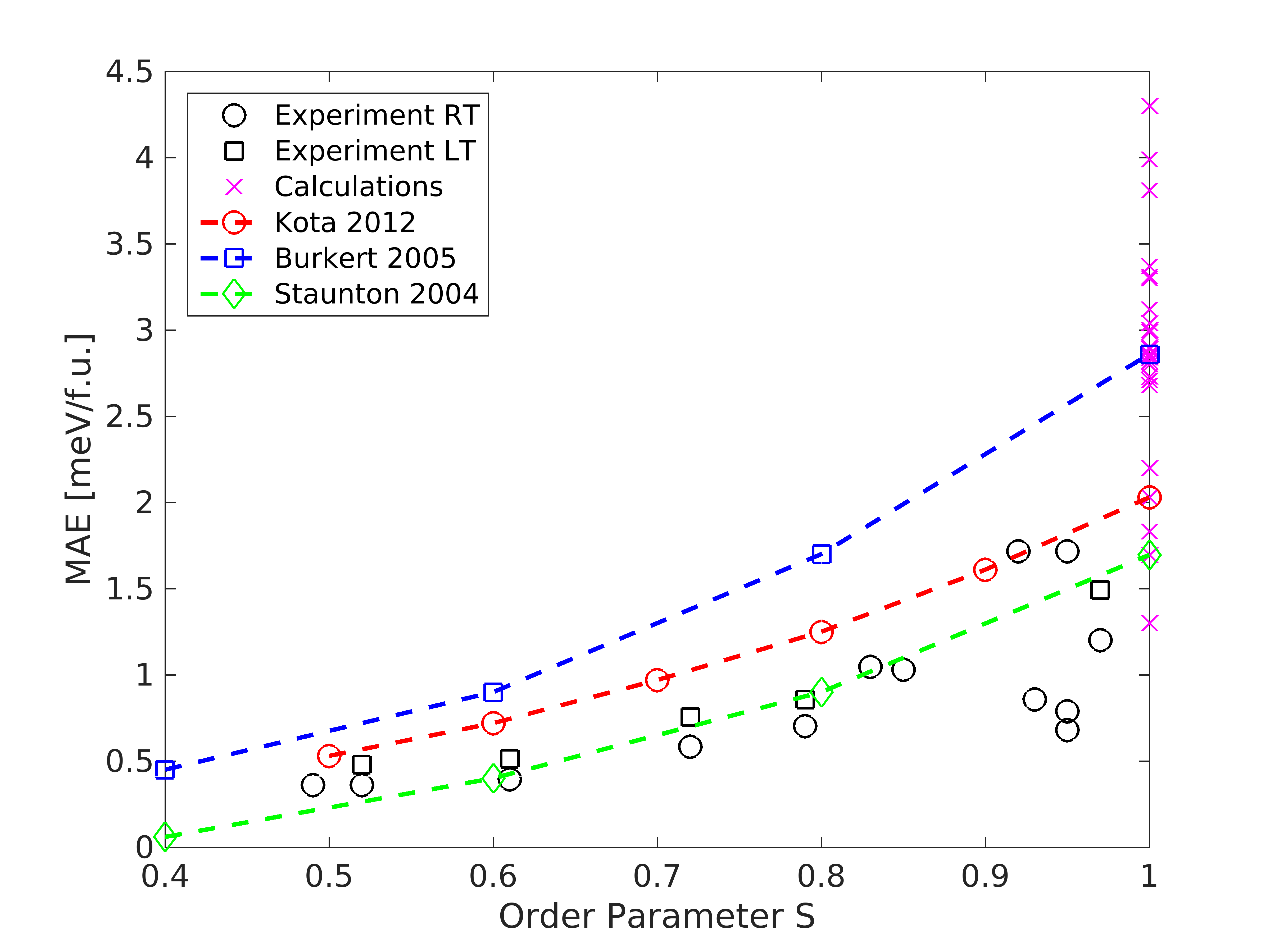}
    \caption{MAE in meV/f.u. plotted over the long range order parameter $S$. Purple crosses are the ab-inito results for the fully ordered system as presented also in Fig.\ref{fig:histogram}. Black symbols represent experiments at room temperature (circles) and around \unit[10]{K} (squares). The red circles, blue squares, and green diamonds represent the calculations by Kota~\cite{kota:12}, Burkert~\cite{burkert:05}, and Staunton~\cite{staunton:04a}.}
	\label{fig:Exp_vs_Calc}
\end{figure}

Computational studies investigating the MAE of disordered FePt in the L1$_0$ structure have been conducted by Staunton~\cite{staunton:04a}, Burkert~\cite{burkert:05}, Kota\cite{kota:12} and their respective coworkers \footnote{Ref.~\onlinecite{deak:14} also provides data for disordered structures, but since they do not deviate from Ref.~\onlinecite{staunton:04a}, other than for $S=1$, and two of the authors appear on both papers, we discuss only the earlier reference at this point.}. Several studies also deal with the electronic structure and magnetic properties of the fully disordered alloy in the face centered cubic structure (e.g. Ref.~\onlinecite{khan:17} and the references therein). Generally the coherent potential approximation (CPA) was used in these papers to model the disorder effects on a mean field basis. While the results calculated by Ref.~\onlinecite{staunton:04a} and Ref.~\onlinecite{kota:12} fit the experimental data very well (see Fig.~\ref{fig:Exp_vs_Calc}), they arrive at considerably lower MAE values for the fully ordered system than the majority of other calculations and the new benchmark study by Khan and coworkers~\cite{khan:16}. Furthermore, at certain order parameters some experimental data points lie higher than the CPA calculations, which seems unlikely given that surface effects, grain boundaries and varying grain orientations in experimental samples will likely decrease the MAE compared to the infinite crystal size of the calculations. Burkert \textit{et al.}~report data that agree with the benchmarks for full order and approach the experimental data nicely for lower values of $S$ (see Fig.~\ref{fig:Exp_vs_Calc}). However, the mean field approach is unable to predict which types of defects are responsible for the significant drop of $K_u$ with decreasing order and the divergence between the studies by Staunton, Kota, and Burkert, all using very similar methods, is a little unsatisfactory.

\begin{figure}[htbp]
	\centering
    \includegraphics[width=0.75\linewidth]{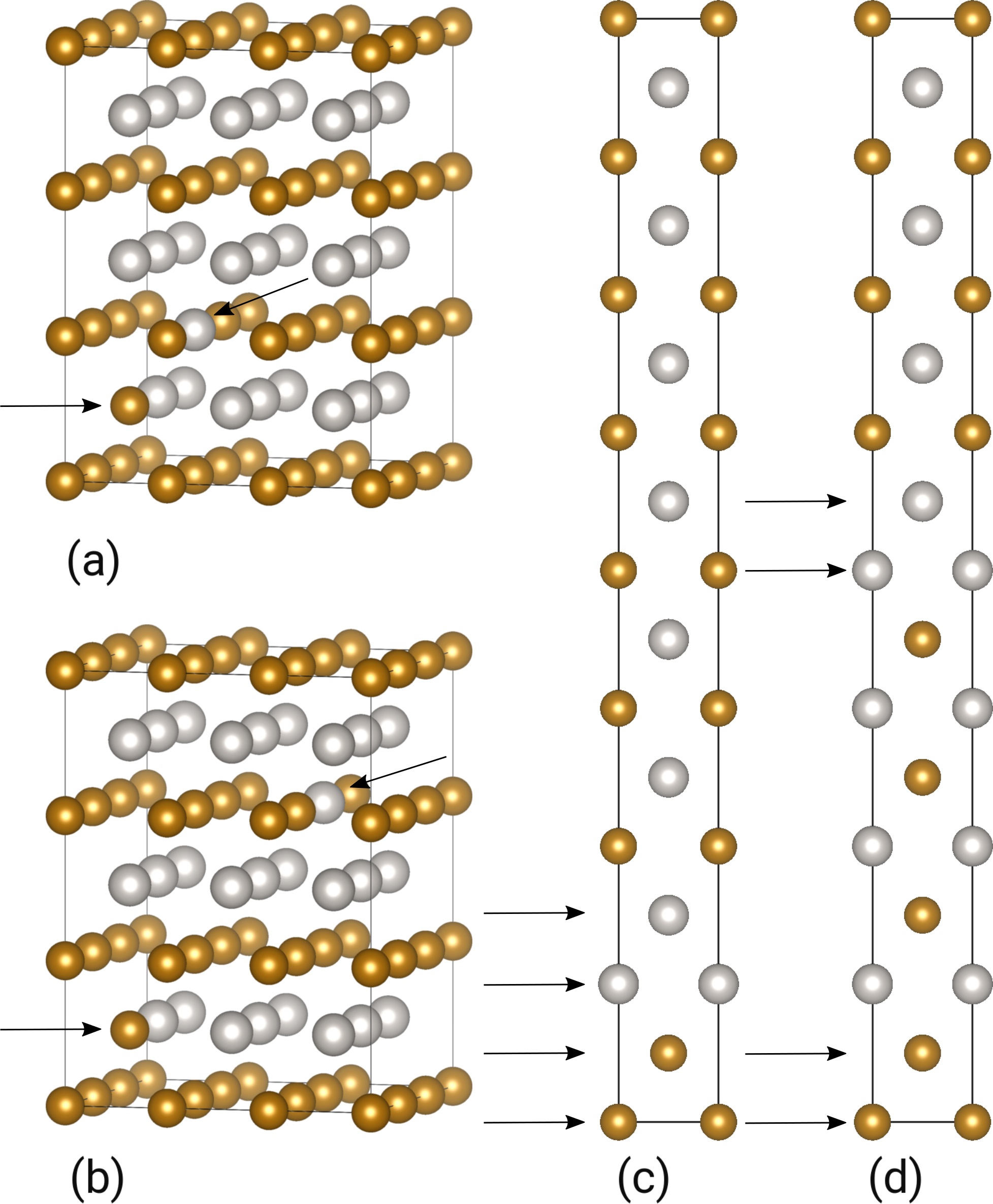}
    \caption{Depiction of a localized (a) and a dispersed (b) anti-site defect, as well as a localized (c) and a dispersed (d) stacking fault in stoichiometric FePt alloy. ASDs are modeled in a 54 atom and SFS in a 16 atom supercell, with Fe shown in gold and Pt in silver. Arrows mark the defect positions.}
	\label{fig:Cells}
\end{figure}

We, on the other hand, are more interested in the influence of single localized defects in the FePt crystal, especially anti-site defects (ASD) and stacking faults (SF). An ASD consists of one Fe and one Pt atom exchanging their place in the lattice, while a SF can occur during growth of FePt thin films if instead of perfect alternating stacking of Fe and Pt planes, two planes of the same material follow each other. We distinguish between localized defects, where two neighboring atoms are exchanged for ASDs and two layers of one type are followed by two layers of the other in SF, and dispersed defects, where the exchanged atoms and the double-planes are far away from each other. These basic defects are depicted in Fig.~\ref{fig:Cells}.

\subsubsection{Defect formation energies}
\label{subsub:defects}

While we do not consider a change in cell volume, the atomic positions in all of our supercells have been relaxed carefully and separate static calculations are used to determine the defect formation energies. As we only consider defects where two (or more) atoms exchange their positions and keep the alloy fully stoichiometric, the defect formation energy (DFE) is simply the total energy of the supercell containing the defect minus $n$ times the total energy of a fully relaxed FePt unit cell, where $n$ is the number of FePt pairs in the supercell, $E_\mathrm{df}=E_\mathrm{sc}-n E_\mathrm{uc}$. Defect formation energies for different super cell sizes (described as multiples of the unit cell in $a$, $b$, and $c$ direction) are given in table~\ref{tab:DFE} for both ASDs and SFs. If 2 defects are considered in a cell the DFEs are averaged over several configurations. For example if 2 local ASDs are put into a $2 \times 2 \times 2$ supercell with the first one being located at the origin, the second on can be shifted by a lattice vector $a$ (equivalent to a shift by $b$), a lattice vector $c$, by both $a$ and $c$ (equivalent to $b$ and $c$), $a$ and $b$, or by $a$, $b$, and $c$ together. We thus arrive at 5 different possibilities, of which two have to be counted twice since they have less symmetry. The DFEs are actually quite different, ranging from \unit[435]{meV} for stacking along $c$, to \unit[690]{meV} for stacking along $a$ or $b$.

\begin{table}[hbt]
\caption{Defect formation energies $E_\mathrm{df}$ per defect for ASDs and SFs with corresponding order parameters $S$ in different supercells and configurations. $N$ is equal to the number of defects per cell, while L or C denote a local or a dispersed defect.}
\label{tab:DFE}
\begin{ruledtabular}
\begin{tabular}{lcccc}
 \noalign{\vskip 1mm}
\multicolumn{5}{c}{ASDs} \\
\noalign{\vskip 1mm}
& $N$ & L/D & $S$ & $E_\mathrm{df}$ [meV] \\
\hline
\noalign{\vskip 1mm}
$4 \times 4 \times 4$ & 1 & L & 0.97 & 735.6 \\
$3 \times 3 \times 3$ & 1 & L & 0.93 & 781.1 \\
$3 \times 3 \times 3$ & 1 & D & 0.93 & 938.5 \\
$3 \times 3 \times 3$ & 2 & L & 0.85 & 714.6 \\
$2 \times 2 \times 2$ & 1 & L & 0.75 & 745.5 \\
$2 \times 2 \times 2$ & 2 & L & 0.50 & 566.9 \\
 \noalign{\vskip 1mm}
\multicolumn{5}{c}{SFs} \\
\noalign{\vskip 1mm}
& $N$ & L/D & $S$ & $E_\mathrm{df}$ [meV] \\
\hline
\noalign{\vskip 1mm}
$1 \times 1 \times 12$ & 1 & L & 0.83 & 449.4 \\
$1 \times 1 \times 10$ & 1 & L & 0.80 & 451.6\\
$1 \times 1 \times 10$ & 1 & D & 0.80 & 455.8 \\
$1 \times 1 \times 8$ & 1 & L & 0.75 & 451.6 \\
$1 \times 1 \times 8$ & 1 & D & 0.75 & 453.5 \\
$1 \times 1 \times 6$ & 1 & L & 0.66 & 449.4 \\
$1 \times 1 \times 5$ & 1 & L & 0.60 & 450.2 \\
$1 \times 1 \times 4$ & 1 & L & 0.50 & 453.9 \\
$1 \times 1 \times 3$ & 1 & L & 0.33 & 447.1 \\
$1 \times 1 \times 2$ & 1 & L & 0.00 & 443.2 \\
\end{tabular}
\end{ruledtabular}
\end{table}

We immediately notice that SFs have a lower DFE than ASDs, and that they are very well decoupled from each other, since the energy stays nearly constant at \unit[$\sim 450$]{meV}. If the stacking fault is localized, with 2 Pt layers followed immediately by two Fe layers, or dispersed, where the double layers are far away from each other does not matter much from an energetic point of view. On the other hand, a single local ASD shows quite different $E_\mathrm{df}$ depending on supercell size. For the $3 \times 3 \times 3$ supercell the DFE is noticeable higher than for the largest cell considered, but the $2 \times 2 \times 2$ shows again a reduced $E_\mathrm{df}$. This indicates that ASDs are not decoupled, and interact attractively in close proximity, as can be seen from the averaged DFE for two ASDs in a $2 \times 2 \times 2$ supercell, which, at \unit[$\sim 566$]{meV} per defect, is considerably lower than an isolated ASD (\unit[$\sim 736$]{meV}).

From the data of table~\ref{tab:DFE} we see that the DFEs are quite sizable at \unit[$\sim 0.5$]{eV} to \unit[0.9]{eV} per defect, depending on type and configuration. Thus, healing out defects in a FePt alloy with low order parameter $S$ would lower the free energy enormously, even considering the decreased entropy. The difficulties in producing highly ordered films of the material experimentally (see section~\ref{sec:Intro}), leads thus to the conclusion that the barriers for defect healing must be comparably large.

\subsubsection{Magnetic anisotropy energy}
\label{subsub:MAE}

\begin{figure}[htbp]
	\centering
    \includegraphics[width=\linewidth]{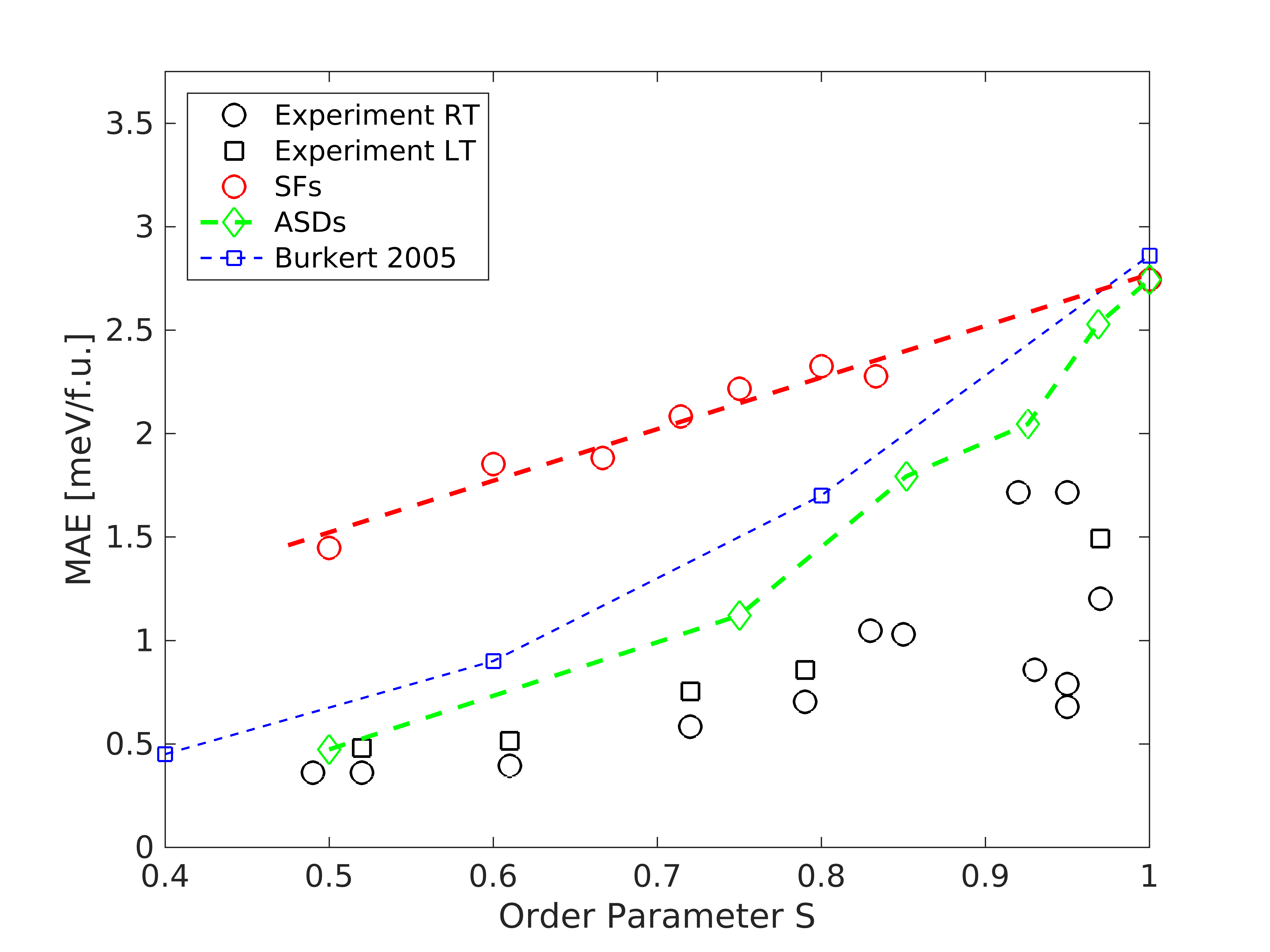}
    \caption{MAE in meV/f.u. plotted over the long range order parameter $S$. Black symbols represent experiments at room temperature (circles) and around \unit[10]{K} (squares). The red circles are for SFs, green diamonds are for ASDs, and blue squares are the KKR-CPA results of Ref.~\onlinecite{burkert:05}. SF data are fitted with a straight line, all other lines serve only as guides to the eye.}
	\label{fig:MAE}
\end{figure}

The magnetic anisotropy energy was calculated for the supercells of table~\ref{tab:DFE} analogous to the method used for the perfect crystal with the same energy cutoff and k-grid density. Our results for both SFs and ASDs are plotted in Fig.~\ref{fig:MAE} alongside the mean field data from Burkert \textit{et al.}~\cite{burkert:05} and the experimental data detailed in Fig.~\ref{fig:Exp}.

It is obvious that although the DFE of ASDs is higher than for SFs, the former are responsible for the strong decrease of the MAE at reduced order.
For stacking faults a decrease in $S$ leads to a linear reduction of the MAE. This means that the defects are not only well isolated from each other regarding the DFE, but also regarding the MAE. As the fraction of correctly ordered unit cells in the supercell decreases, the MAE decreases proportionally, right down $S=0.5$. If the cell size is further reduced to six atoms ($S=0.33$) or 4 atoms ($S=0$), which are not shown in Fig.~\ref{fig:MAE}, it is not really appropriate to speak of a SF, as 2/3 or more of the material is layered in the wrong way. For $S=0.33$ we calculate an MAE of \unit[1.40]{meV}, slightly higher than the linear trend would predict, while for $S=0$ the MAE drops to \unit[0.18]{meV}. 
An anti site defect on the other hand, has much larger effect, which can also be reasoned intuitively, as there are significantly more unit cells directly influenced by a single localized ASD (8) than by a SF (2), disregarding lower order effects like atoms sitting on wrong lattice sites in a neighboring unit cell or relaxations. Furthermore ASDs perturb the surrounding of Pt atoms (which are mainly responsible for the large MAE) in 3D, while SFs only change the surrounding in 2D, having a diminished effect on the MAE. While our calculations for ASDs are in reasonable agreement with the experimental data over the whole range, the agreement is certainly a lot better for lower values of $S$. This might indicate that ASDs do indeed cluster together in FePt,  as our data for $S=0.5$ are averaged over different configurations of 2 ASDs in an 16 atom supercell. This result is also supported by the DFE data in table~\ref{tab:DFE}.

The KKR-CPA results from Burkert \textit{et al.}~\cite{burkert:05} lies between our SF and ASD data, although closer to the ASD points. This is to be expected from a mean field approach for random disorder, which should encounter SF like regions less than ASDs. Delocalized defects where not included in Fig.~\ref{fig:MAE}, due to generally higher DFE, but the cases that we tested showed MAE differing less than 5\% from the localized defects in the same super cell.

In the single-atom resolution images from Ref.~\onlinecite{yang:17} ASDs are also commonly observed and their density is still $\sim 3\%$ in the highest ordered grain centers of the nanoparticle~\footnote{In the non stoichiometric sample of Ref.~\onlinecite{yang:17} a swap defect denotes what we call here ASD. An anti site defect in their notation is just a single Fe atom sitting on a Pt site or vice versa.}. This is a strong indication that ASDs are also common in stoichiometric FePt, although they have a rather large DFE. Although we can not completely rule out that the MAE might also be lowered slightly by correlation effects (see section~\ref{subsub:RPA}), we believe that the inclusion of ASDs is sufficient to explain the experimental MAE data on the basis of standard DFT calculations.

%% file: Conclusion.tex
We have shown that ASDs are responsible for the large reduction of the MAE in FePt with decreasing long range order parameter $S$. Experimental measurements and ab-initio DFT calculations are thus also comparable without including many body effects beyond the LDA or GGA level. Qualitative calculations using the ACDFT-RPA show that the effect of more accurate treatment of correlations is probably smaller than that of disorder. This will allow future DFT calculations to accurately model thin FePt films and layered systems useful for magnetic recording with reasonable effort.

%% file: Acknowledgements.tex
The authors would like to thank Josef Redinger and Herwig Michor for fruitful discussions, and acknowledge the support by the Austrian Science Fund (FWF)
[SFB ViCoM F4109-N28 and F4112-N28]. The ample support of computer resources by the Vienna Scientific Cluster
(VSC) are gratefully acknowledged. Fig.~\ref{fig:Cells} in this paper was created with help of the VESTA code~\cite{vesta:11}.